\documentclass[twocolumn,showpacs,prb,amsmath,amssymb,floatfix]{revtex4}

\usepackage{dcolumn}
\usepackage{amsmath,amssymb}
\usepackage{bm}
\usepackage{chngpage}
\usepackage{epsfig}
\usepackage{slashed}
\usepackage{dsfont}

\newcommand{\nn}{\nonumber}

\begin{document}
\title{Renormalized transport properties of randomly gapped 2D Dirac fermions}
\author{Andreas Sinner and Klaus Ziegler}
\affiliation{Institut f\"ur Physik, Universit\"at Augsburg}
\date{\today}

\begin{abstract}
We investigate the scaling properties of the recently acquired fermionic non--linear $\sigma$--model which controls gapless
diffusive modes in a two--dimensional disordered  system of Dirac electrons beyond charge neutrality.  The transport on large
scales is governed by a renormalizable nonlocal field theory.  For zero mean random gap, it is characterized
by the absence of a dynamic gap generation and a scale invariant diffusion coefficient. The $\beta$ function of the
DC conductivity, computed for this model, is in perfect agreement with numerical results obtained previously.  
\end{abstract}
\maketitle

Transport in systems whose band structure has a node structure (e.g. Dirac points), as it appears in graphene and on the
surface of 3D topological insulators, has been the subject of intense research recently.
The experimental observation of transport in graphene is characterized by a minimal conductivity
at the charge neutrality (or Dirac) point and by a linearly increasing conductivity with increasing 
(electron or hole) charge density \cite{novoselov05,zhang05}. 
Thus, experimentally it is easy to distinguish the very robust minimal conductivity
which is contrasted by the disorder-dependent conductivity away from the Dirac point. The behavior at the Dirac point 
has also been predicted by field theory, showing that the minimal conductivity is quite independent (or very weakly dependent)
on disorder. For instance, a nonlinear sigma model approximation \cite{Ziegler1997,Ziegler2009,Ziegler2012} 
as well as perturbation theory in terms of disorder strength clearly 
shows a very weak disorder dependence \cite{Sinner2011}.
This has led to the claim that ballistic transport cannot be distinguished from diffusive 
transport at the Dirac point. However, transport properties away from the Dirac point are theoretically not easily accessible.
There are several attempts, based on a classical Boltzmann approach, which predict the experimentally
observed linearly increasing conductivity as we go away from the Dirac point. However, it was only recently
that a more general field-theoretical approach, based on the Kubo formalism, was suggested to describe
the transport of 2D Dirac  fermions within a unified theory \cite{Ziegler2012,Ziegler2012a}, using a four-body Hamiltonian.

In this paper we focus on disorder due to a random gap. This case is particularly interesting because it
can lead to a metal-insulator transition when the average gap is equal to a critical value \cite{Ziegler2009}.
Starting from a nonlinear sigma model that controls the diffusive modes, we study  the renormalization
of the interaction of these modes as well as the renormalization of the diffusion coefficient and the conductivity
at the Dirac point and away from it.

\section{Model} 

Below we give a brief sketch of the field theoretical approach to the conductivity. Here we are led by the representation given 
in Refs.~[\onlinecite{Ziegler2012,Ziegler1997,Ziegler2009,Bocquet2000}]. The main quantity to be computed is the disorder averaged two--particle 
Green's function. The disorder potential $v$ of the strength $g$ is supposed to have zero mean $\langle v^{}_r \rangle =  0$ and 
Gaussian correlator $\langle v^{}_r v^{}_{r^\prime} \rangle = g\delta^{}_{rr^\prime}$. For random gap disorder, the disorder averaged two--particle Green's function reads
\begin{eqnarray}
\nn
&\displaystyle
K^{}_{rr^\prime}  = -\langle{\rm Tr}^{}_n[G^{}_{rr^\prime}(i\epsilon)\sigma^{}_1 G^{\rm T}_{r^\prime r}(i\epsilon)\sigma^{}_1]\rangle^{}_v
&
\\
&\displaystyle
= \sum_{m,m^\prime,n,n^\prime} [\sigma^{}_1]^{}_{mn}[\sigma^{}_1]^{}_{n^\prime m^\prime}\langle\phi^1_{r^\prime m^\prime}\bar\phi^1_{rm}\phi^2_{rn} \bar\phi^2_{r^\prime n^\prime}\rangle^{}_{\phi},
&
\end{eqnarray}
where Tr$^{}_n$ is taken on the extended Dirac space, $\phi$ is a four component superfield $\phi=(\psi^{}_{1,+},\psi^{}_{1,-},\chi^{}_{2,-},\chi^{}_{2,+})$, consisting of 
a complex $\psi^\ast_{1,\pm}$ and a Grassmann $\chi^\ast_{2,\pm}$ field. In our notation $\sigma^{}_{1,2,3}$ are usual Pauli matrices and $\sigma^{}_0$ the $2\times2$ unity matrix. 
The field averaging is defined as 
\begin{equation}
\langle \cdots \rangle^{}_\phi =  \int{\cal D}[\phi]\cdots e^{-{\cal S}},
\end{equation}
with the action
\begin{equation}
\label{eq:AvAct}
{\cal S} = -i(\phi\cdot(\bar H^{}_0 + i\bar\epsilon)\bar\phi) + 
g(\phi \cdot \bar\sigma^{}_3\bar\phi)^2,
\end{equation}
where $\bar\sigma^{}_3=1^{}_4\otimes\sigma^{}_3$, $\bar\epsilon=\epsilon 1^{}_4\otimes\sigma^{}_0$, $1^{}_4$ the  $4\times4$ unity matrix, and 
\begin{eqnarray}
\label{eq:StHam}
\bar H^{}= 
\left(
\begin{array}{cccc}
 H^{} +  \mu & 0 & 0 & 0 \\
 0 &  H^{} -\mu & 0 & 0 \\
 0 &  0 & H^{\rm T} - \mu & 0 \\
 0 &  0 &  0 & H^{\rm T} + \mu
\end{array}
\right),
\end{eqnarray}
with the chemical potential $\mu$, the non-random Hamiltonian $H^{}_0=i\sigma\cdot\nabla$
and the random Hamiltonian 
\begin{equation}
H = H^{}_0 + v \sigma^{}_3. 
\end{equation}
Then $\bar H^{}$ is invariant under the global symmetry transformation 
\begin{equation}
\bar H = e^{\bar S} \bar H e^{\bar S},
\end{equation}
where $\bar S$ is given by the following matrix 
\begin{equation}
\label{eq:SimiTr}
\bar S = 
\left(
\begin{array}{cccc} 
 0 & 0 & \varphi^{}_1 \sigma^{}_1 & 0 \\
 0 & 0 &   0 & \varphi^{}_2 \sigma^{}_1 \\
\varphi^\prime_1 \sigma^{}_1 & 0 & 0 & 0 \\
 0  & \varphi^\prime_2 \sigma^{}_1 & 0 & 0
\end{array}
\right),
\end{equation}
with two scalar  fields $\varphi^{}_1$ and $\varphi^{}_2$, which obey Grassmann statistics, i.e. $\varphi^{}_i\varphi^\prime_i = - \varphi^\prime_i\varphi^{}_i$ and $\varphi^{}_i\varphi^{}_j = - \varphi^{}_j\varphi{}_i$. 

We decouple the interaction term in Eq.~(\ref{eq:AvAct})  by a Hubbard--Stratonovich transformation. Integrating out superfields $\phi$ yields an action    in terms of composite supersymmetric Hubbard--Stratonovich fields $\bar Q$
\begin{equation}
\label{eq:HSTrAct}
{\cal S}^\prime = \frac{1}{g}{\rm Trg}(\bar Q)^2 + \log{\rm detg}(\bar H^{}_0 +i\bar\epsilon + 2\bar Q\bar \sigma^{}_3).
\end{equation}
A non--trivial vacuum of this theory $\bar Q^{}_0$ is found from the saddle--point condition and turns out to be degenerated with respect to the  transformation $\bar S$ defined in Eq.~(\ref{eq:SimiTr}):
\begin{equation}
e^{\bar S}\bar Q^{}_0 e^{-\bar S} = \bar Q^{}_1 + \bar Q^{}_2 e^{-2\bar S},
\end{equation}
where $\bar Q^{}_1$ ($\bar Q^{}_2$) commutes (anticommutes) with $\bar S$, and vanishes under the graded trace ${\rm Trg}Q^{}_0 = 0$. On the saddle--point manifold, the action represents a fermionic non--linear $\sigma$--model~[\onlinecite{Ziegler2012}] 
\begin{equation}
\label{eq:NLSM1}
{\cal S}^\prime = \log{\rm detg}(\bar H^{}_0 +i\bar\epsilon + 2\bar Q^{}_1\sigma^{}_3 +2 \bar Q^{}_2 \bar \sigma^{}_3 e^{2\bar S}).
\end{equation}
The field $\bar Q^{}_2$ represents the order parameter for the spontaneous breaking of the symmetry generated by $\bar S$. Expanding Eq.~(\ref{eq:NLSM1}) up to second order in $\bar Q^{}_2$ as 
\begin{equation}
{\cal S}^\prime = {\cal S}^{}_0 + {\cal S}^{\prime\prime}, 
\end{equation}
and using the exact relation for Grassmann fields
\begin{equation} 
e^{2\bar S} = 1 + 2\bar S + 2\bar S^2
\end{equation}
yields 
\begin{eqnarray}
\nn
&\displaystyle
{\cal S}^{\prime\prime} = 4{\rm Trg}\left[ 
\bar G^{}_0 \bar Q^{}_2 \bar \sigma^{}_3 \bar S^2 \right. &\\
\label{eq:NLSM}
&\displaystyle\left.
+ 2(\bar G^{}_0 \bar Q^{}_2 \bar \sigma^{}_3 \bar S)^2 + 2(\bar G^{}_0 \bar Q^{}_2 \bar \sigma^{}_3 \bar S^2)^2
\right], &
\end{eqnarray}
with matrix Green's function 
\begin{equation}
\bar G^{}_{0,rr^\prime} = {\rm diag}\left\{g^{}_+,g^{}_-,g^{\rm T}_-,g^{\rm T}_+\right\}^{}_{rr^\prime},
\end{equation}
and
\begin{equation}
g^{}_{\pm,rr^\prime} = [H ^{}_0 + i(\epsilon + \eta \pm  i\mu)\sigma^{}_0 ]^{-1}_{rr^\prime},
\end{equation}
where $\eta\sim\exp[-\pi/g]$ is the scattering rate~\cite{Ziegler1997,Ziegler2009}. Eventually, we rewrite Eq.~(\ref{eq:NLSM}) in terms of scalar Grassmann fields $\varphi^{}_j$ and obtain a nonlocal fermionic theory  
\begin{eqnarray}
\nn
&\displaystyle 
{\cal S}[\varphi] = \frac{g\eta}{2}\sum_{j=1,2}\sum_{rr^\prime}\left[\varphi^{}_{jr^\prime} \delta^{}_{r^\prime r} (i\epsilon-D\nabla^2)\varphi^\prime_{jr} \right.
&\\
\label{eq:InitFullAct} 
&\displaystyle \left.- 2\eta^2(-1)^j
\sum_{s=\pm}{\rm Tr}^{}_2\left\{sg^{}_{s,rr^\prime} g^{}_{s,r^\prime r}\right\} \varphi^{}_{jr^\prime}\varphi^\prime_{jr^\prime}\varphi^{}_{jr}\varphi^\prime_{jr}\right],
&
\hspace{3mm}
\end{eqnarray}
which describes the diffusion of Dirac electrons. For $\mu~<~\eta$, the diffusion coefficient reads
\begin{equation}
\label{eq:DiffCoeff}
D \approx \frac{1}{2\pi \eta} + {\cal O}(\mu).
\end{equation}

\section{Renormalization group analysis} 

Below we investigate the scaling properties of action Eq.~(\ref{eq:InitFullAct}) at large distances. 
For this purpose we expand the Fourier transform of the vertex function $\displaystyle \sum_{s=\pm}{\rm Tr}^{}_2\left\{sg^{}_{s,rr^\prime} g^{}_{s,r^\prime r}\right\}$ to the leading order in momenta of the fields. This changes action Eq.~(\ref{eq:InitFullAct}) to
\begin{eqnarray}
\label{eq:InitAct}
\nn
&\displaystyle
{\cal S}[\varphi] = 
\sum_{j,q}~\varphi^{}_{jq}(D^\prime q^2+i\epsilon^\prime)\varphi^\prime_{jq} &
\\
&\displaystyle 
-i\lambda \sum_{j,kqpt} (-1)^{j}\delta^{}_{k-q,p-t} (p-t)^2 \varphi^{}_{jk}\varphi^\prime_{jq}\varphi^{}_{jp}\varphi^\prime_{jt}, &
\;\;\;
\end{eqnarray}
with the shorthands $\sum_q=\int d^2q/(2\pi)^2$, $\delta_{k,p}=(2\pi)^2\delta(k+p)$, $\epsilon^\prime=g\eta\epsilon/2$, and $D^\prime = g\eta D/2$. The interaction strength is defined as
\begin{equation}
\label{eq:IntStrength} 
\lambda = \frac{2}{3\pi}\frac{\mu\eta^3}{(\eta^2+\mu^2)^2}\approx\frac{2}{3\pi}\frac{\mu}{\eta}.
\end{equation}
The zeroth order term in field momenta of the interaction part is zero due to its locality, while the first order term vanishes by the symmetry. The frequency $\epsilon$ is supposed to be small and sent to zero in the DC limit. For this reason we do not distinguish between $\epsilon^\prime$ and $\epsilon$.

In order to find the infrared behavior of action Eq.~(\ref{eq:InitAct}) we follow the usual prescription of the Wilson RG transformation~\cite{Shankar1994}:  We decompose Grassmann fields into fast $\varphi^{}_f$ and slow $\varphi^{}_s$ modes. The idea is to integrate out fast modes and to obtain an action which mimics action Eq.~(\ref{eq:InitAct}) but contains solely slow fields. To the second order in DC perturbation theory this action reads
\begin{eqnarray}
\nn
&\displaystyle
\bar {\cal S}[\varphi^{}_s] \approx {\cal S}^{}_0[\varphi^{}_s] + 
{\cal S}^{}_{\rm int}[\varphi^{}_s] +  \langle {\cal S}^{}_{\rm int}[\varphi^{}_s,\varphi^{}_f]\rangle^{DC}_{f} 
&\\
\label{eq:NewAction}
&\displaystyle
- \frac{1}{2} \langle {\cal S}^{}_{\rm int}[\varphi^{}_s,\varphi^{}_f]{\cal S}^{}_{\rm int}[\varphi^{}_s,\varphi^{}_f]\rangle^{DC}_{f},
&
\end{eqnarray}
with
\begin{eqnarray}
\nn
{\cal S}^{}_{\rm int}[\varphi^{}_s] &=& \displaystyle 
-i\lambda \sum_{j,kqpt} (-1)^{j} \delta^{}_{k-q,p-t} (p-t)^2 \\
&&\displaystyle
\varphi^{}_{sjk}\varphi^\prime_{sjq}\varphi^{}_{sjp}\varphi^\prime_{sjt},
\label{eq:IntAct}
\end{eqnarray}
and ${\cal S}^{}_{\rm int}[\varphi^{}_s,\varphi^{}_f]$ represents terms which contain both slow and fast fields. The averaging operator reads
\begin{equation}
\langle\cdots\rangle^{DC}_{f} =\frac{1}{{\cal Z}^{DC}_0} \int~{\cal D}[\varphi^{}_f]~\cdots~e^{-{\cal S}^{DC}_0[\varphi^{}_f]},
\end{equation}
with the free DC action
\begin{equation}
{\cal S}^{DC}_0[\varphi^{}_f] = D^\prime \sum_{j,r}\nabla\varphi^{}_{jrf} \nabla\varphi^\prime_{jrf}.
\end{equation}
Obviously, this construction guarantees $\langle 1 \rangle^{DC}_{f} = 1$ and defines the DC propagator (in Fourier representation)
\begin{equation}
\langle \varphi^{}_{iqf}\varphi^\prime_{jkf} \rangle^{}_{f} = \frac{\delta^{}_{ij}\delta^{}_{q,-k}}{D^\prime q^2}.
\end{equation}

\begin{figure}[t]
\includegraphics[height=1.5cm]{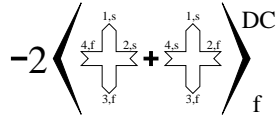}
\caption{Diagrams responsible for renormalization of $\epsilon$. }
\label{fig:eps} 
\end{figure}

\begin{figure}[t]
\includegraphics[height=1.5cm]{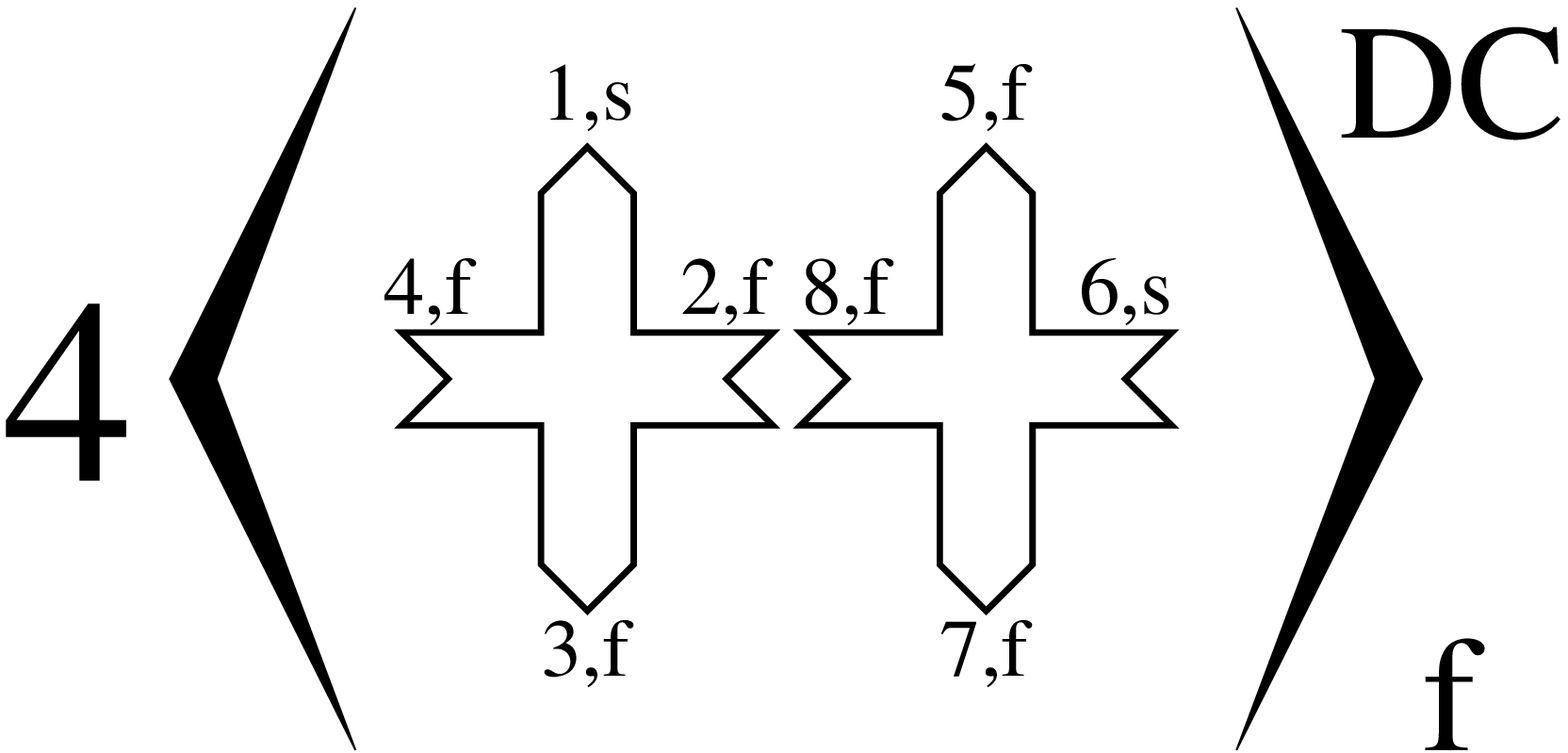}
\caption{Diagrams responsible for renormalization of $D^\prime$. }
\label{fig:diff} 
\end{figure}

\begin{figure}[t]
\includegraphics[height=1.5cm]{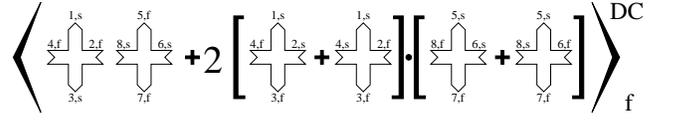}
\caption{Diagrams responsible for renormalization of $\lambda$. }
\label{fig:lambda} 
\end{figure}

The derivation of the renormalization group equations for the frequency $\epsilon$, 'diffusion' coefficient $D^\prime$ and interaction strength $\lambda$ of the action Eq.~(\ref{eq:InitAct}) is a challenging task. Diagrams which have to be evaluated arise by merging vertices shown in Figs.~  \ref{fig:eps},~\ref{fig:diff} and~\ref{fig:lambda}. The renormalization of the energy $\epsilon$ comes from one--loop diagrams which emerge by averaging vertices depicted in Fig.~\ref{fig:eps} over fast fields. As shown in Appendix~\ref{app:eps}, the evaluation of the diagrams yields a result that does not develop any divergences in the infrared, since the vertex is proportional to the squared loop momentum, and the propagator to inverse squared loop momentum:
\begin{equation}
\label{eq:FreqRen}
\displaystyle \bar\epsilon = \epsilon + \frac{(-1)^j}{2\pi}\frac{\lambda}{D^\prime }\Lambda^2_0,
\end{equation}
where $\Lambda_0$ denotes an upper cutoff. This expression is not a renormalization group equation in the strict sense, since it does not contain the running cutoff parameter $\ell=\log\Lambda^{}_0/\Lambda$. It represents a kind of a finite size effect which disappears in the continuous limit.  

The renormalization of the diffusion coefficient can be obtained by integrating out fast fields in vertices depicted in Fig.~\ref{fig:diff}. 
The evaluation of the functional integral is presented in Appendix~\ref{app:DiffCoeff} and yields the following renormalization of the free action:
\begin{equation}
\label{eq:gam1}
\sum_{i,p} \gamma(p)\varphi^{}_{ip}\varphi^\prime_{ip} \approx \sum_{i,r}\left[\bar D^\prime \nabla\varphi^{}_{ir}\nabla\varphi^\prime_{ir} + m \varphi^{}_{ir}\varphi^\prime_{ir}\right],
\end{equation}
where $m$ and $\bar D^\prime$ are expansion coefficients of zeroth and second order in momentum $p$ of the vertex function
\begin{equation}
\label{eq:gam2}
\gamma(p) = 4\frac{(i\lambda)^2}{{D^\prime}^3} \sum^{}_{kq}~\frac{(k+p)^2[(k+p)^2-(q+p)^2]}{k^2 q^2 (k+q+p)^2}. 
\end{equation}
There is no linear term in this expansion, since $\gamma(p)$ is symmetric with respect to the sign mirroring of $p$, i.e. $\gamma(-p)=\gamma(p)$. In the Appendices~\ref{app:mass} and~\ref{app:diff} is shown that both coefficients $m$ and $\bar D^\prime$ vanish. This is a very important result, since it guarantees the preservation of the gapless diffusive mode and the realty of the diffusion coefficient even for the complex interaction strength. Therefore, the only running parameter is the interaction strength $\lambda$. Its renormalization is due to the one--loop diagrams which emerge after integrating out fast fields in Fig.~\ref{fig:lambda}. Lengthy and elaborate calculations presented in Appendix~\ref{app:lambda} lead to the remarkably simple renormalization group equation 
\begin{equation}
\label{eq:RGInt} 
\partial^{}_\ell (i\lambda^{}_j) = \frac{(-1)^j}{\pi}\frac{(i\lambda^{}_j)^2}{{D^\prime}^2},
\end{equation}
for each fermionic channel. This equation is easily solved with the same starting value $\lambda^{}_{j0}=u^{}_0$ in both channels, $u^{}_0$ given in Eq.~(\ref{eq:IntStrength})
\begin{eqnarray}
\label{eq:SolLambda} 
\lambda^{}_j = \frac{u^{}_0}{\displaystyle 1+\frac{u^2_0\ell^2}{\pi^2 D^{\prime4}}}+
i\frac{(-1)^j}{\pi D^{\prime2}}\frac{u^2_0\ell}{\displaystyle 1+\frac{u^2_0\ell^2}{\pi^2 D^{\prime4}}}.
\end{eqnarray}

\begin{figure}[t]
\includegraphics[width=7cm]{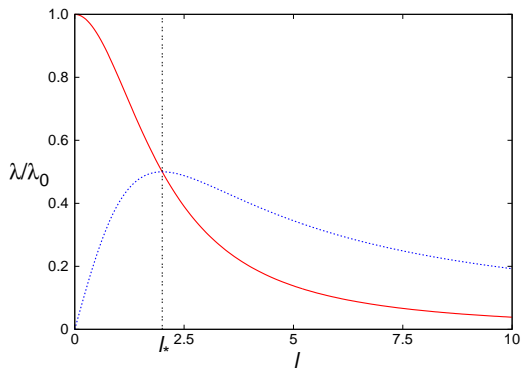}
\caption{(Color online) Renormalization of the interaction strength $\lambda^{}_2$. Solid (red) line shows the real part and dashed 
(blue) line  the imaginary part of $\lambda^{}_2$. The cross--over scale is shown by the vertical (black) dotted line at $\ell^{}_\ast = \pi D^{\prime2}u^{-1}_0$.}
\label{fig:RGFlow}
\end{figure}

\begin{figure}[t]
\includegraphics[width=8cm]{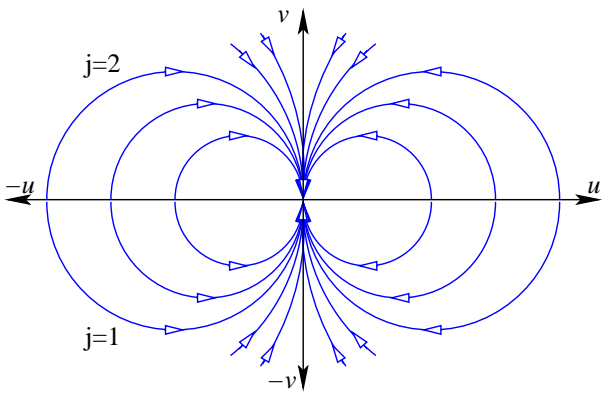}
\caption{(Color online) Renormalization group flow of the interaction $\lambda^{}_j$ in the parametric space spanned by 
$u = {\Re}\lambda^{}_j$ and $v={\Im}\lambda^{}_j$. The flow in the upper halfplane corresponds to the Grassmann field $\varphi^{}_2$, that in the lower plane to $\varphi^{}_1$.}
\label{fig:RGFParam}
\end{figure}

Both real and imaginary parts of the interaction scale down to zero but not equally fast. 
At large scales, the imaginary part of $\lambda$ becomes dominant and therefore generates a genuine real interaction. 
The RG flow for the Grassmann field $\varphi^{}_2$ is depicted in Fig.~\ref{fig:RGFlow}. Fig.~\ref{fig:RGFParam} 
shows the RG landscape in the parametric space spanned by the real and imaginary part of the interaction $\lambda^{}_j$. 
The RG trajectories represent a set of excentric circles with diameter $u^{}_0$, each attracted to the Gaussian fixed point 
at $\Re\lambda^{}_j=0$ and $\Im\lambda^{}_j=0$. This can be seen best if we substitute $\lambda^{}_j = u^{}_j + i v^{}_j$ in Eq.~(\ref{eq:RGInt}). Then we get
\begin{equation}
\left\{
\begin{array}{ccl}
\partial^{}_\ell u^{}_j &=& (-1)^{1+j}~2u^{}_j v^{}_j,\\
\\
\partial^{}_\ell v^{}_j &=& (-1)^j~(u^2_j - v^2_j).
\end{array}
\right.
\end{equation}
The right hand side of this system of differential equations represents indeed a parametrized circle.

Independently from the choice of the initial value both real and imaginary parts of $\lambda^{}_2$ (analogously for $\lambda^{}_1$) become equal at the length obtained from the condition:
$$
\frac{u^{}_0 \ell^{}_\ast}{\pi D^{\prime2}} = 1,\;\;{\rm with}\;\, \ell^{}_\ast = \log\frac{\xi}{l},
$$
with $l$ denoting the mean free path, which gives 
\begin{equation}
\label{eq:DelLenMain}
\xi = l \exp\left[\frac{\pi {D^\prime}^2}{u^{}_0}\right] \approx l\exp\left[\frac{3g^2}{32}\frac{\eta}{\mu}\right]\;\;{\rm for}\;\;\mu<\eta.
\end{equation}
Here we used definitions of the bare diffusion coefficient Eq.~(\ref{eq:DiffCoeff}) and  interaction strength Eq.~(\ref{eq:IntStrength}). At half filling, i.e. for $\mu=0$, this scale is infinite. 


\section{Scaling properties of the DC conductivity} 

Our ultimate task is to determine the scaling behavior of the 
DC conductivity. For this we need to compute the corrections to the conductivity which arise due to the doping. 
The DC conductivity is either determined from the Einstein relation
$\bar \sigma\propto \rho D$ ($\rho$ is the density of states at the Fermi level) or calculated from the Kubo formula,
\begin{equation}
\label{eq:Kubo}
\bar \sigma = 2 \epsilon^2 \left.\frac{\partial}{\partial q^2} \bar K(q)\right|_{q=0}, 
\end{equation}
where the two-particles Green's function takes contributions from both channels $j=1,2$ into account:
\begin{equation}
\label{eq:2PGF}
\bar K(q) = \frac{1}{g}\sum_{ij,p} ~ \langle \varphi^{}_{iq}\varphi^\prime_{jp} \rangle.
\end{equation}
Here, the functional integral should be performed over the full action Eq.~(\ref{eq:InitAct}): 
\begin{equation}
\label{eq:AvOp}
\langle\cdots\rangle = \frac{1}{\cal Z} 
\int{\cal D}[{\varphi}] \cdots e^{-{\cal S}[\varphi]},
\end{equation}
with $\langle1\rangle=1$. To the leading order in $\lambda^{}_j$, the two--particles Green's function is approximated  as
\begin{equation}
\label{eq:PT1ord}
\bar K(q) \approx \frac{1}{g}\sum^{}_{ij,p}\left(\langle \varphi^{}_{iq}\varphi^\prime_{jp} \rangle^0 - 
\langle \varphi^{}_{iq}\varphi^\prime_{jp}{\cal S}^{}_{\rm int} \rangle^0\right),
\end{equation}
where ${\cal S}^{}_{\rm int}$ is given in Eq.~(\ref{eq:InitAct}). The functional integration is to be performed over the free action only. As shown in Appendix~\ref{app:cond}, we eventually arrive at the following one--loop RG equation for the conductivity
\begin{equation}
\label{eq:CondRgEq}
\partial^{}_\ell\sigma = -2i\frac{\sigma^{}_0}{gD^\prime} \sum_{j=1,2}\left[(-1)^j\lambda^{}_j\right].
\end{equation}
Further progress can be made if we exploit Eq.~(\ref{eq:SolLambda}):
\begin{equation}
\partial^{}_\ell\sigma 
= \frac{4\sigma^{}_0}{\pi g D^{\prime3}} \frac{u^2_0\ell}{\displaystyle 1+\frac{u^2_0\ell^2}{\pi^2 D^{\prime4}}}.
\end{equation}
Unpleasant constants can be eliminated by rescaling $u^{}_0\to\pi D^{\prime2}u^{}_0$ and using definition of the diffusion coefficient $D^\prime=g/4\pi$. This finally yields 
\begin{equation}
\label{eq:CondScale}
\partial^{}_\ell \sigma = \sigma^{}_0 \frac{u^2_0\ell}{1+u^2_0\ell^2}.
\end{equation}
The integration of this equation is simple and we obtain the following asymptotic expression for the conductivity 
\begin{equation}
\label{eq:RunCond} 
\sigma(u^{}_0\ell) = \sigma^{}_0 + \frac{\sigma^{}_0}{2} \log[1+u^2_0\ell^2].
\end{equation}
At half filling, i.e. $u^{}_0=0$, the conductivity does not flow, i.e. it is scale invariant. At large scales, 
i.e. for $L\gg\xi$, $\xi$ given in~(\ref{eq:DelLenMain}), the conductivity grows bi-logarithmically as function 
of the sample size $\sigma(L)\sim\sigma^{}_0\log\log L/\xi$. For this reason $\xi$ can be associated with an intermediate
localization scale. Due to the infrared asymptotic freedom of the underlying model, this result should be asymptotically correct in all loops. 

\begin{figure}[t]
\includegraphics[width=7cm]{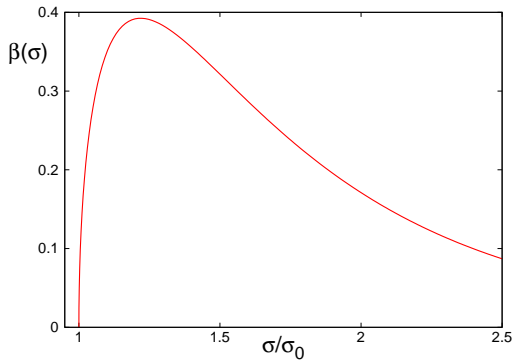}
\caption{(Color online) $\beta$ function corresponding to the conductivity in Eq.~(\ref{eq:RunCond}).}
\label{fig:BetaFunc}
\end{figure}

\section{Discussion and conclusions} 

The scaling properties of the conductivity are usually given by the $\beta$ function
\begin{equation}
\beta(\sigma) = \frac{d}{dl}\log\sigma,
\end{equation}
with rescaled logarithmic length $l=u^{}_0\ell$. The conductivity from Eq.~(\ref{eq:RunCond}) generates the
$\beta$ function as depicted in Fig.~\ref{fig:BetaFunc}. Its shape reveals a striking resemblance of the numerically determined
$\beta$ function of graphene at the Dirac point with random scalar potential disorder~\cite{Bardarson2007}: It starts at
the value of the universal minimal conductivity; it is strictly positive; it reveals a distinct maximum related to the length $\xi$;
it does not have any fixed points besides $\sigma = \sigma^{}_0$ and $\sigma=\infty$. Finally, it does not depend on any
quantities apart from the conductivity $\sigma$ itself, in line with the one--parameter scaling hypothesis. However, one 
might wonder whether the above mentioned bi--logarithmic asymptotics of the conductivity compares well with the predicted
logarithmic growth at the Dirac point~\cite{Bocquet2000,Abergel2010,Bardarson2007}. It is indeed not difficult to reproduce
the  $\beta$ function of Ref.~[\onlinecite{Bardarson2007}] by applying the Wilson RG transformation directly to the two--particle
Green's function and exploiting the scaling properties of the disorder strength $g$. Instead, in our approach we keep $g$ 
scale invariant. This assumption suits well for weak disorder, provided the sample size is much smaller than disorder 
generated intrinsic length $\sim\exp[1/g]$~\cite{Schuessler2009}, which corresponds to the common experimental situation.
Under these circumstances the scale invariance of the conductivity was demonstrated both numerically and
analytically~\cite{Bardarson2010,Sinner2011}. On the  other hand, in finite samples the conductivity grows logarithmically 
as function of the chemical potential: $\sigma(u^{}_0)\sim\sigma^{}_0\log(\mu/\eta)$ in accordance with
Ref.~[\onlinecite{Aleiner2006}].

In conclusion, we have presented a scaling analysis of the diffusion coefficient and the DC conductivity of doped graphene 
with a random gap. For this purpose we have used an alternative field theory and investigated its scaling properties. 
On this basis we derived an invariant diffusion coefficient and an astonishingly simple expressions for the scaling of 
the conductivity that reproduces the distinct shape of the $\beta$ function of disordered graphene, found previously in
numerical calculations~\cite{Bardarson2007}. 

\section*{ACKNOWLEDGMENTS}
 We acknowledge financial support by the DFG grant ZI 305/5--1.

\appendix

\section{Renormalization of the energy $\epsilon$}
\label{app:eps}

For sake of simplicity we use below the following notation for the interaction part of the action Eq.~(\ref{eq:InitAct}):
\begin{eqnarray}
\nn
{\cal S}^{}_{\rm int} &=& \displaystyle -i\lambda \sum_j (-1)^j\int d1d2d3d4~ \delta(1+3-2-4)\\
\nn 
&& \displaystyle 
\times(3-4)^2 \varphi^{}_{j1}\varphi^\prime_{j2}\varphi^{}_{j3}\varphi^\prime_{j4}. 
\end{eqnarray}
The renormalization of $\epsilon$ is due to diagrams which are obtained by contracting fast fields in Fig.~\ref{fig:eps}. Contracting is possible in the only way, i.e. the diagrams have one--fold degeneracy. To take Grassmann statistics correctly into account it is necessary to permute fast fields through, such that they form 'normal ordered' pairs $\langle\varphi\varphi^\prime\rangle$: 
\begin{eqnarray}
\nn
&\displaystyle
2i\lambda\sum_{j}(-1)^j\int d1d2d3d4 ~(3-4)^2 \delta(1+3-2-4)&\\
\nn
&\displaystyle
\left[\varphi^{}_{j1s} \varphi^\prime_{j2s} \dot\varphi^{}_{j3f} \dot\varphi^\prime_{j4f} + \varphi^{}_{j1s} \dot\varphi^\prime_{j2f}  \dot\varphi^{}_{j3f} \varphi^\prime_{j4s} \right]&\\
\nn
&\displaystyle
= 2i\lambda \sum_{j}(-1)^j\int d1d2d3d4 ~(3-4)^2 \delta(1+3-2-4)&\\
\nn
&\displaystyle
\left[\varphi^{}_{j1s} \varphi^\prime_{j2s} \langle\dot\varphi^{}_{j3f} \dot\varphi^\prime_{j4f}\rangle - \varphi^{}_{j1s}\varphi^\prime_{j4s}\langle\dot\varphi^{}_{j3f} \dot\varphi^\prime_{j2f}\rangle    \right].&
\end{eqnarray}
The fields to be contracted are marked with black dotes.  Then, the contractions can be performed and we obtain:
\begin{eqnarray}
\nn
&\displaystyle
2i\lambda \sum_{j}(-1)^j\int d1d2d3d4 ~(3-4)^2 \delta(1+3-2-4)&\\
\nn
&\displaystyle
\left[\varphi^{}_{j1s} \varphi^\prime_{j2s} \delta(3-4)\Pi(4) - \varphi^{}_{j1s} \varphi^\prime_{j4s}\delta(2-3)\Pi(3) \right],& 
\end{eqnarray}
where $\Pi(q) = {1}/{D^\prime q^2}$. First contribution is zero because
\begin{equation}
\nn
\int d3~ (3-4)^2\delta(3-4) = 0. 
\end{equation}
Second contribution is finite and cutoff dependent: 
\begin{eqnarray}
\nn
&\displaystyle
-2i\lambda\int d3~ \Pi(3) \sum_{j} (-1)^j \int d1~(3-1)^2 \varphi^{}_{j1s}\varphi^\prime_{j1s}
&\\
\nn
&\displaystyle
\approx -2i\lambda\int d3~ \Pi(3) (3)^2 
\sum_{j} (-1)^j \int d1~\varphi^{}_{j1s}\varphi^\prime_{j1s}.&
\end{eqnarray}
The renormalization factor then reads:
\begin{equation}
\nn
-2i\lambda\int d3~ \Pi(3) (3)^2  = -\frac{i}{2\pi} \frac{\lambda}{D^\prime }(\Lambda^{2}_0 - \Lambda^2),
\end{equation}
which reduces for $\Lambda\to0$ to
$-i\lambda\Lambda^2_0/2\pi D^\prime$. Lifting it into the exponent and absorbing into the action gives Eq.~(\ref{eq:FreqRen}).

\section{Renormalization of the diffusion coefficient}
\label{app:DiffCoeff}

The renormalization of the diffusion coefficient comes from the diagrams constructed from vertices depicted in Fig~\ref{fig:diff}. Every diagram is twice degenerated, i.e. the analytical expression reads

\begin{eqnarray}
\nn
&\displaystyle
4(-i\lambda)^2 \sum_{ij}(-1)^{i+j} \int d1d2d3d4 \int d5d6d7d8 &\\
\nn
& (3-4)^2(7-8)^2 ~ \delta(1+3-2-4)\delta(5+7-6-8) &\\
\nn
&\displaystyle
\times
\left[ 
\varphi^{}_{s1i}\dddot\varphi^\prime_{f2i}\ddot\varphi^{}_{f3i}\dot\varphi^\prime_{f4i}\dddot\varphi^{}_{f5j}\varphi^\prime_{s6j}\dot\varphi^{}_{f7j}\ddot\varphi^\prime_{f8j} \right.&\\
\nn
&\displaystyle
\left. 
+\varphi^{}_{s1i}\dddot\varphi^\prime_{f2i}\ddot\varphi^{}_{f3i}\dot\varphi^\prime_{f4i}\dot\varphi^{}_{f5j}\varphi^\prime_{s6j}\dddot\varphi^{}_{f7j}\ddot\varphi^\prime_{f8j} 
\right].&
\end{eqnarray}
We permute fields through and perform functional integrations in order to get
\begin{eqnarray}
\nn
&\displaystyle
 4(i\lambda)^2 \sum_{ij}(-1)^{i+j} \int d1d2d3d4 \int d5d6d7d8 ~\varphi^{}_{1si}\varphi^\prime_{6sj}&\\
\nn
&\displaystyle 
\times(3-4)^2(7-8)^2 \delta(1+3-2-4)\delta(5+7-6-8) &\\
\nn
&\displaystyle \times
\left[
\langle\varphi^{}_{f5j}\varphi^\prime_{f2i}\rangle \langle\varphi^{}_{f3i}\varphi^\prime_{f8j}\rangle \langle\varphi^{}_{f7j}\varphi^\prime_{f4i}\rangle \right.&\\
\nn
&\displaystyle
\left. 
- \langle\varphi^{}_{f7j}\varphi^\prime_{f2i}\rangle \langle\varphi^{}_{f3i}\varphi^\prime_{f8j}\rangle \langle\varphi^{}_{f5j}\varphi^\prime_{f4i}\rangle
\right]&
\\
\nn
&\displaystyle 
= 4(i\lambda)^2 \sum_{ij}(-1)^{i+j} \delta^{}_{ij}\delta^{}_{ij}\delta^{}_{ij} \int d1d2d3d4 \int d5d6d7d8 &\\
\nn
&\displaystyle 
\times(3-4)^2(7-8)^2 \delta(1+3-2-4)\delta(5+7-6-8) 
&\\
\nn
&\displaystyle
 \varphi^{}_{1si}\varphi^\prime_{6sj} ~ \Pi(2)\Pi(4)\Pi(8) 
[\delta(5-2)\delta(3-8)\delta(7-4) &\\
\nn
& - \delta(7-2)\delta(3-8)\delta(5-4)] &\\
\nn
&\displaystyle = 4(i\lambda)^2 \sum_i \int d1~\varphi^{}_{1i}  \varphi^\prime_{1i}\int d2d4~(2-1)^2 &\\
\nn 
&\displaystyle \times [(2-1)^2-(4-1)^2]\Pi(2)\Pi(4)\Pi(2+4-1) &\\
\nn
&\displaystyle = \sum_{i,p} \gamma(p)\varphi^{}_{ip}\varphi^\prime_{ip}.&
\end{eqnarray}
The function $\gamma(p)$ is defined in Eqs.~(\ref{eq:gam1}) and~(\ref{eq:gam2}). The way, how momenta in arguments of $\delta$--functions are integrated out, is not unique. Therefore we can shift integration variables when necessary.

\subsection{Computation of the mass term}
\label{app:mass}

First we evaluate the expression for the mass:
\begin{eqnarray}
\nn
&\displaystyle
m \propto \int\frac{d^2k}{(2\pi)^2}\int\frac{d^2q}{(2\pi)^2}\frac{k^2[k^2-q^2]}{k^2 q^2 (k+q)^2} &\\
\nn
&\displaystyle 
= \int\frac{d^2k}{(2\pi)^2}\int\frac{d^2q}{(2\pi)^2} 
\left[ 
\frac{k^2}{q^2(k+q)^2} - \frac{1}{(k+q)^2}
\right].&
\end{eqnarray}
Shifting in the second term $q \to q-k$ we have
\begin{equation}
\nn
\int\frac{d^2k}{(2\pi)^2}\int\frac{d^2q}{(2\pi)^2} \frac{1}{q^2} = \frac{\ell}{2\pi}\int\frac{d^2k}{(2\pi)^2},
\end{equation}
where $\ell = \log {\Lambda^{}_0}/{\Lambda}$. First term is conveniently evaluated using Feynman parametrization:
\begin{eqnarray}
\nn
&\displaystyle
\int\frac{d^2k}{(2\pi)^2}\int\frac{d^2q}{(2\pi)^2} \frac{k^2}{q^2(k+q)^2} = \int\frac{d^2k}{(2\pi)^2} k^2&\\
\nn
&\displaystyle
 \times \int_0^1 dx \int\frac{d^2q}{(2\pi)^2} \frac{1}{[(1-x)q^2+x(k+q)^2]^2} = (\ast).&
\end{eqnarray}
Performing shift $q\to q-xk$ we get
\begin{eqnarray}
\nn
&\displaystyle
(\ast)=\int\frac{d^2k}{(2\pi)^2} k^2 \int_0^1 dx \int\frac{d^2q}{(2\pi)^2} \frac{1}{[q^2+x(1-x)k^2]^2} &\\
\nn
&\displaystyle
= \frac{1}{4\pi} \int\frac{d^2k}{(2\pi)^2} k^2 \int_0^1 dx ~ \frac{1}{k^2x(1-x)} 
&\\
\nn
&\displaystyle
= \frac{1}{4\pi}\int_0^1 dx \left[\frac{1}{x}+\frac{1}{1-x}\right] \int\frac{d^2k}{(2\pi)^2} &\\
\nn
&\displaystyle
= \frac{1}{2\pi} \int^1_{e^{-\ell}} \frac{dx}{x} \int\frac{d^2k}{(2\pi)^2} = \frac{\ell}{2\pi}\int\frac{d^2k}{(2\pi)^2},
&
\end{eqnarray}
i.e. the very same result. Therefore, the mass is zero, the diffusive Goldstone mode is preserved.

\subsection{Computation of the diffusion coefficient renormalization}
\label{app:diff}

Next we evaluate the renormalization of the diffusion coefficient:
\begin{eqnarray}
\nn
&\displaystyle
\bar D^\prime  = 4\frac{(i\lambda)^2}{{D^\prime}^3} \frac{1}{2}
\int\frac{d^2q}{(2\pi)^2}\int\frac{d^2k}{(2\pi)^2}
&\\
\nn
&\displaystyle
\left.\frac{\partial^2}{\partial p^2}\left[\frac{(q+p)^4}{k^2q^2(k+q+p)^2} -\frac{(q+p)^2(k+p)^2}{k^2q^2(k+p+q)^2}\right]
\right|_{p=0}.
& \;\;\;
\end{eqnarray}
We start with first term. Using Feynman parametrization we have
\begin{eqnarray}
\nn
&\displaystyle
{\rm I}=\left.\frac{\partial^2}{\partial p^2}
\int\frac{d^2q}{(2\pi)^2}\int\frac{d^2k}{(2\pi)^2}~\frac{(q+p)^4}{k^2q^2(k+q+p)^2} 
\right|_{p=0}
&\\
\nn
&\displaystyle
= \frac{\partial^2}{\partial p^2}
\int\frac{d^2q}{(2\pi)^2} \frac{(q+p)^4}{q^2}
\int_0^1 dx &\\
\nn
&\displaystyle 
\left.
\times \int\frac{d^2k}{(2\pi)^2}~\frac{1}{[(1-x)k^2+x(k+q+p)^2]^2}
\right|_{p=0}.
&
\end{eqnarray}
Next shift $k\to k-x(q+p)$:
\begin{eqnarray}
\nn
&\displaystyle
\frac{\partial^2}{\partial p^2}
\int\frac{d^2q}{(2\pi)^2} \frac{(q+p)^4}{q^2}
\int_0^1 dx & \\
\nn
&\displaystyle
\left.\times\int\frac{d^2k}{(2\pi)^2}~\frac{1}{[k^2+x(1-x)(q+p)^2]^2}\right|_{p=0}  
&\\
\nn
&\displaystyle
=\frac{1}{4\pi}\left.\frac{\partial^2}{\partial p^2}\int\frac{d^2q}{(2\pi)^2} \frac{(q+p)^4}{q^2}
\int_0^1 dx~ \frac{1}{(q+p)^2x(1-x)}\right|_{p=0} &\\
\nn
&\displaystyle
 = \frac{1}{4\pi}\left.\frac{\partial^2}{\partial p^2}\int\frac{d^2q}{(2\pi)^2} \frac{(q+p)^2}{q^2}\right|_{p=0}
\int_0^1 dx~ \frac{1}{x(1-x)} 
&\\
&\displaystyle
\nn
= \frac{1}{4\pi}\int\frac{d^2q}{(2\pi)^2} \frac{2}{q^2}\int_0^1 dx~ \frac{1}{x(1-x)} = 2\left(\frac{\ell}{2\pi}\right)^2.
&
\end{eqnarray}
It is important to recognize that shifting of the integration variables with respect to the external momentum $p$ does not affect the final result. This is because the integration over the momentum conserving $\delta$--function is not unique. Indeed, if we shift $q\to q-p$ in the above integral we obtain the same result:
\begin{eqnarray}
\nn
&\displaystyle
\left.\frac{\partial^2}{\partial p^2}
\int\frac{d^2q}{(2\pi)^2}\frac{q^4}{(q-p)^2}\int\frac{d^2k}{(2\pi)^2}~\frac{1}{k^2(k+q)^2}\right|_{p=0} &\\
\nn
&\displaystyle
= \frac{\ell}{2\pi}\left.\frac{\partial^2}{\partial p^2}\int\frac{d^2q}{(2\pi)^2}\frac{q^2}{(q-p)^2}\right|_{p=0} ,&
\end{eqnarray}
where we skipped integration over the momentum $k$. After mirroring sign of $q$ we have 
\begin{eqnarray}
\nn
&\displaystyle
\frac{\ell}{2\pi}\left.\frac{\partial^2}{\partial p^2}  \int\frac{d^2q}{(2\pi)^2}\frac{q^2}{(q+p)^2} \right|_{p=0} &\\
\nn
&\displaystyle
= \frac{\ell}{2\pi} \int\frac{d^2q}{(2\pi)^2}~\left[8 \frac{(\hat e_p\cdot q)^2}{q^4} - \frac{2}{q^2}\right] = (\ast). &
\end{eqnarray}
Here we have to shed some light on the structure of the first term:
\begin{eqnarray}
\nn
&\displaystyle
\int\frac{d^2q}{(2\pi)^2} \frac{(\hat e_p\cdot q)^2}{q^4}= \int^{\Lambda^{}_0}_\Lambda 
\frac{qdq}{(2\pi)^2}\frac{1}{q^4}\int_0^{2\pi} d\alpha (\hat e_p\cdot q)^2
&\\
\nn
&\displaystyle
= \int^{\Lambda^{}_0}_\Lambda \frac{qdq}{(2\pi)^2} \frac{1}{q^4} \int_0^{2\pi} d\alpha q^2 \cos^2(\alpha-\varphi)
&\\
\nn
&\displaystyle
=  \int^{\Lambda^{}_0}_\Lambda \frac{qdq}{(2\pi)^2} \frac{1}{q^4} \int_0^{2\pi} d\alpha  \frac{q^2}{2}(1 + \cos 2(\alpha-\varphi))
&\\
\nn
&\displaystyle
=  \int^{\Lambda^{}_0}_\Lambda \frac{qdq}{(2\pi)^2} \frac{1}{2q^2} \int_0^{2\pi} d\alpha  =  \int\frac{d^2q}{(2\pi)^2} \frac{1}{2q^2}.
&
\end{eqnarray}
Thus we continue:
\begin{eqnarray}
\nn
&\displaystyle
(\ast) =  \frac{\ell}{2\pi} \int\frac{d^2q}{(2\pi)^2}~\left[\frac{4}{q^2} - \frac{2}{q^2}\right] 
&\\
\nn
&\displaystyle
= 2\frac{\ell}{2\pi}\int\frac{d^2q}{(2\pi)^2}\frac{1}{q^2}=2\left(\frac{\ell}{2\pi}\right)^2,
&
\end{eqnarray}
i.e. the very same result as before. This knowledge can now be used for evaluating second term:
\begin{eqnarray}
\nn
{\rm II} = \left.\frac{\partial^2}{\partial p^2}
\int\frac{d^2q}{(2\pi)^2}\int\frac{d^2k}{(2\pi)^2}~\frac{(q+p)^2(k+p)^2}{k^2q^2(k+p+q)^2}\right|_{p=0}.
\end{eqnarray}
Here we are allowed to shift $k\to k-p$, which yields 
\begin{eqnarray}
\nn
&\displaystyle
\int\frac{d^2q}{(2\pi)^2}\int\frac{d^2k}{(2\pi)^2}~\frac{k^2}{q^2(k+q)^2} \left.\frac{\partial^2}{\partial p^2}
\frac{(q+p)^2}{(k-p)^2}
\right|_{p=0}
&\\
\nn
&\displaystyle
= \int\frac{d^2q}{(2\pi)^2}\int\frac{d^2k}{(2\pi)^2}~\frac{k^2}{q^2(k+q)^2} \left[\frac{2}{k^2} \right.&\\
\nn
&\displaystyle
\left.
+ q^2\left(8\frac{(\hat e_p\cdot k)^2}{k^6} - \frac{2}{k^4}\right) + 8\frac{(\hat e_p\cdot q)(\hat e_p\cdot k)}{k^4}
\right]
&\\
\nn
&\displaystyle
= \int\frac{d^2q}{(2\pi)^2}\int\frac{d^2k}{(2\pi)^2}~\frac{2}{q^2(k+q)^2} &\\
\nn
&\displaystyle
+ \int\frac{d^2q}{(2\pi)^2}\int\frac{d^2k}{(2\pi)^2}~\frac{1}{(k+q)^2}\left[8\frac{(\hat e_p\cdot k)^2}{k^4} - \frac{2}{k^2}\right]
&\\
\nn
&\displaystyle
+ 8 \int\frac{d^2q}{(2\pi)^2}\int\frac{d^2k}{(2\pi)^2} \frac{(\hat e_p\cdot q)(\hat e_p\cdot k)}{q^2k^2(k+q)^2} .
&
\end{eqnarray}
First and second terms can be easily evaluated after shifting $k\to k-q$ and $q\to q-k$, respectively. Both give the same contribution ${\rm I}$ calculated above. The last term is more cumbersome:
\begin{eqnarray} 
\nn
&\displaystyle
8 \int\frac{d^2q}{(2\pi)^2}\int\frac{d^2k}{(2\pi)^2} \frac{(\hat e_p\cdot q)(\hat e_p\cdot k)}{q^2k^2(k+q)^2} = 8 \int\frac{d^2q}{(2\pi)^2}\frac{\hat e_p\cdot q}{q^2}&\\
\nn
&\displaystyle
\times\int_0^1 dx~\int\frac{d^2k}{(2\pi)^2} \frac{\hat e_p\cdot k}{[(1-x)k^2+x(k+q)^2]^2} .& 
\end{eqnarray}
Again, we shift $k\to k-x q$, which gives 
\begin{eqnarray} 
\nn
&\displaystyle
\int\frac{d^2q}{(2\pi)^2}\frac{\hat e_p\cdot q}{q^2}\int_0^1 dx 
\int\frac{d^2k}{(2\pi)^2} \frac{8\hat e_p\cdot (k-xq)}{[k^2+x(1-x)q^2]^2} =&\\
\nn
&\displaystyle
-\int\frac{d^2q}{(2\pi)^2}\frac{(\hat e_p\cdot q)^2}{q^2}\int_0^1 dx  
\int\frac{d^2k}{(2\pi)^2} \frac{8x}{[k^2+x(1-x)q^2]^2} 
&\\
\nn
&\displaystyle
= -8 \int\frac{d^2q}{(2\pi)^2}\frac{(\hat e_p\cdot q)^2}{q^4} \frac{1}{4\pi}\int_0^1 \frac{dx}{1-x} &\\
\nn
&\displaystyle 
= \frac{1}{\pi}\int\frac{d^2q}{(2\pi)^2}\frac{1}{q^2}\int^{e^{-\ell}}_1\frac{dy}{y} = -2\left(\frac{\ell}{2\pi}\right)^2 = -{\rm I}, 
&
\end{eqnarray}
where at some point we have substituted $y=1-x$. Summing over all contributions to ${\rm II}$ we obtain 
$$
{\rm II} = {\rm I} +  {\rm I} - {\rm I} = {\rm I}.
$$
The renormalization of the diffusion coefficient to order $\lambda^2$ is therefore zero: 
$$
\bar D^\prime  \propto {\rm I} - {\rm II} = 0.
$$

\section{Renormalization of the interaction strength}
\label{app:lambda}

The renormalization of the interaction is due to one--loop diagrams which arise after merging vertices depicted in Fig.~\ref{fig:lambda}.
We consider both contributions separately. 

\subsection{Contribution I}
\begin{center}
\includegraphics[height=15mm]{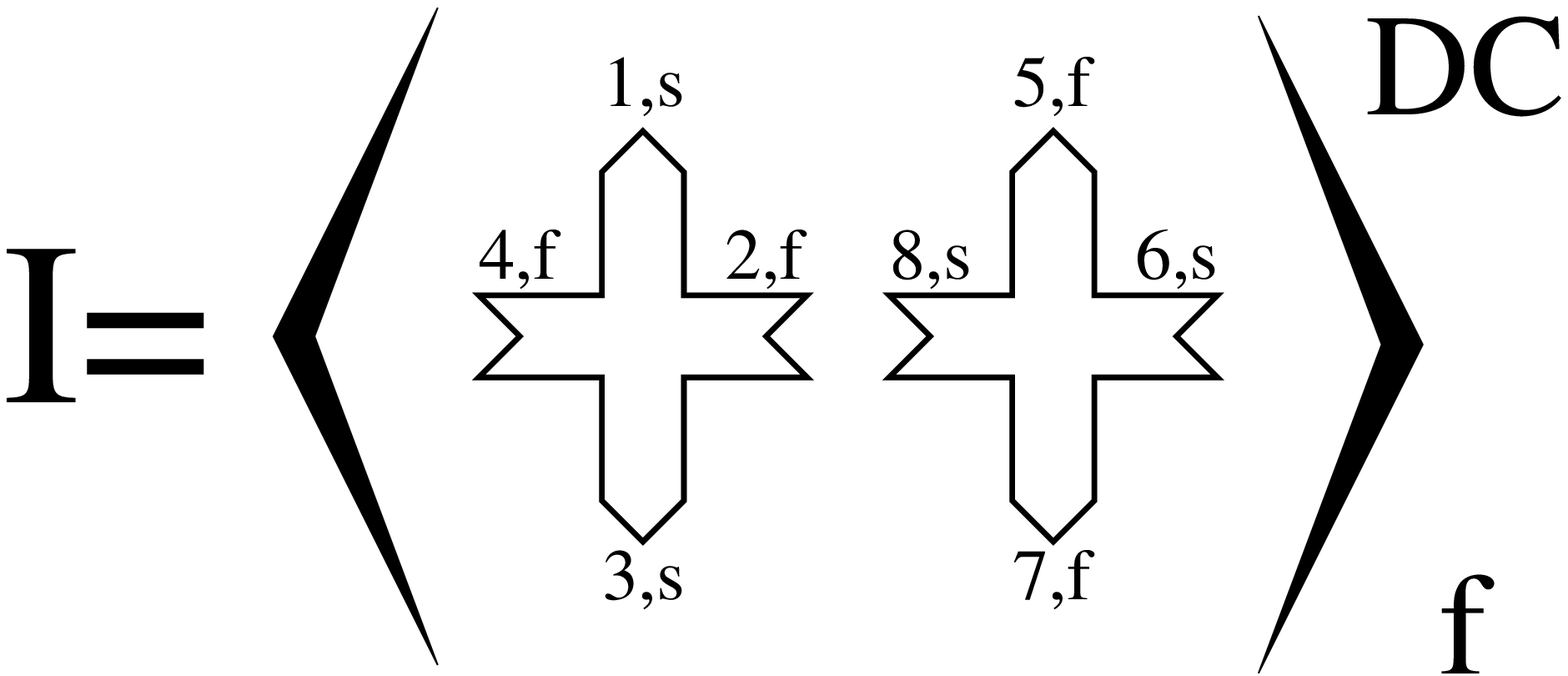}
\end{center}
It has a two--fold degeneration, i.e. there are two possibilities for contracting fast fields: 
\begin{eqnarray}
\nn
&\displaystyle {\rm I} = (-i\lambda)^2 \sum_{ij}(-1)^{i+j} \int d1d2d3d4\int d5d6d7d8 &\\
\nn
&\displaystyle
(3-4)^2(7-8)^2\delta(1+3-2-4)\delta(5+7-6-8) &\\
\nn
&\displaystyle
[\varphi^{}_{s1i} \dot\varphi^\prime_{f2i} \varphi^{}_{s3i} \ddot\varphi^\prime_{f4i} \dot\varphi^{}_{f5j} \varphi^\prime_{s6j} \ddot\varphi^{}_{f7j} \varphi^\prime_{s8j} &\\
&\displaystyle 
+\varphi^{}_{s1i} \ddot\varphi^\prime_{f2i} \varphi^{}_{s3i} \dot\varphi^\prime_{f4i} \dot\varphi^{}_{f5j} \varphi^\prime_{s6j} \ddot\varphi^{}_{f7j} \varphi^\prime_{s8j}].&
\end{eqnarray}
Next, we permute and contract Grassmann fields:
\begin{eqnarray}
\nn
&\displaystyle {\rm I} = (i\lambda)^2 \sum_{ij}(-1)^{i+j} \int d1d2d3d4\int d5d6d7d8 &\\
\nn
&\displaystyle (3-4)^2(7-8)^2\delta(1+3-2-4)\delta(5+7-6-8) &\\
\nn
&\displaystyle
\times\left[ 
\displaystyle (-1)^{3+2} \varphi^{}_{s1i} \varphi^{}_{s3i} \langle\varphi^{}_{f5j}\varphi^\prime_{f2i}\rangle \varphi^\prime_{s6j} \langle\varphi^{}_{f7j}\varphi^\prime_{f4i}\rangle \varphi^\prime_{s8j} \right.&\\
\nn
&\displaystyle
\left. 
+ (-1)^{1+5}\varphi^{}_{s1i} \langle\varphi^{}_{f7j}\varphi^\prime_{f2i}\rangle \varphi^{}_{s3i} \langle\varphi^{}_{f5j}\varphi^\prime_{f4i}\rangle \varphi^\prime_{s6j}  \varphi^\prime_{s8j}
\right]&\\
\nn
&\displaystyle =
(i\lambda)^2 \sum_{ij}(-1)^{i+j} \delta^{}_{ij}\delta^{}_{ji}\int d1d2d3d4\int d5d6d7d8 &\\
\nn
&\displaystyle
(3-4)^2(7-8)^2\delta(1+3-2-4)\delta(5+7-6-8)  &\\
\nn
&\displaystyle
\times\left[\delta(7-2)\delta(5-4) - \delta(5-2)\delta(7-4)\right]&\\
\nn
&\displaystyle
\varphi^{}_{s1i}\varphi^{}_{s3i}\varphi^\prime_{s6j}\varphi^\prime_{s8j}\Pi(2)\Pi(4) .
&
\end{eqnarray}
Performing summations and integrations we arrive at
\begin{eqnarray}
\nn
&\displaystyle
{\rm I} = (i\lambda)^2 \int d1d3d6d8~\delta(1+3-6-8) &\\
\nn
&\displaystyle
\sum_i\varphi^{}_{s1i}\varphi^{}_{s3i}\varphi^\prime_{s6i}\varphi^\prime_{s8i} \int d4~(4-3)^2&\\
\nn
&\displaystyle
\times[(4-6)^2-(4-8)^2]\Pi(4)\Pi(4-3-1).&
\end{eqnarray}
After reordering fields in first term and renaming variables we finally obtain
\begin{eqnarray}
\nn
&\displaystyle
{\rm I} = -2(i\lambda)^2  \int d1d2d3d4\delta(1+3-2-4) &\\
\nn
&\displaystyle
\sum_i\varphi^{}_{i1}\varphi^\prime_{i2}\varphi^{}_{i3}\varphi^\prime_{i4} 
\int d5~(5-3)^2 &\\
\nn
&\displaystyle
\times(5-2)^2\Pi(5)\Pi(5-3-1).&
\end{eqnarray}
Now we expand vertex function up to the second order in fields momenta:
\begin{eqnarray}
\nn
&\displaystyle
\int d5~(5-3)^2(5-2)^2\Pi(5)\Pi(5-3-1) &\\
\nn
&\displaystyle
 = \frac{1}{{D^\prime}^2}\int\frac{d^2q}{(2\pi)^2}~\frac{(q-k)^2(q-p)^2}{q^2(q-k-t)^2} 
&\\
\nn
&\displaystyle
\approx
\left[
kp\frac{\partial^2}{\partial k\partial p} + kt\frac{\partial^2}{\partial k\partial t} + pt\frac{\partial^2}{\partial p\partial t} +
\frac{k^2}{2}\frac{\partial^2}{\partial k^2} + \frac{p^2}{2}\frac{\partial^2}{\partial p^2}\right. &\\
\label{eq:Expans}
&\displaystyle
\left.
 + \frac{t^2}{2}\frac{\partial^2}{\partial t^2}
\right]
\left.\frac{1}{{D^\prime}^2}\int\frac{d^2q}{(2\pi)^2}~\frac{(q-k)^2(q-p)^2}{q^2(q-k-t)^2}\right|_{k,p,t=0}.
&\;\;\;
\end{eqnarray}

\begin{itemize}
\item  Order $kp$: The factor is zero, since
\begin{eqnarray}
\nn
&\displaystyle
\frac{\partial^2}{\partial k\partial p} \left. \int\frac{d^2q}{(2\pi)^2}~\frac{(q-k)^2(q-p)^2}{q^2(q-k)^2}\right|_{k,p=0} &\\
\nn
&\displaystyle
= \frac{\partial^2}{\partial k\partial p} \left. \int\frac{d^2q}{(2\pi)^2}~\frac{(q-p)^2}{q^2}\right|_{k,p=0} = 0
&
\end{eqnarray}
after differentiation with respect to $k$.

\item Order $kt$: The expansion factor is also zero, because:
\begin{eqnarray}
\nn
&\displaystyle
\frac{\partial^2}{\partial k\partial t} \left. \int\frac{d^2q}{(2\pi)^2}~\frac{(q-k)^2 q^2}{q^2(q-k-t)^2}\right|_{k,t=0} &\\
\nn
&\displaystyle
= -\frac{\partial}{\partial k} \left. \int\frac{d^2q}{(2\pi)^2}~\frac{2\hat e_t\cdot(q-k)}{(q-k)^2}\right|_{k=0}
&\\
\nn
&\displaystyle
 = -\int\frac{d^2q}{(2\pi)^2}~\left[4\frac{(\hat e^{}_k\cdot q)(\hat e^{}_t\cdot q)}{q^4} - 2\frac{\hat e_t\cdot \hat e^{}_k}{q^2} \right]\;\;(\ast)
&
\end{eqnarray}
Rewrite first term:
\begin{eqnarray}
\nn
&\displaystyle
\int_0^{2\pi} d\varphi (\hat e^{}_t\cdot q)(\hat e^{}_k\cdot q) &\\
\nn
&\displaystyle
= q^2\int_0^{2\pi} d \varphi \cos(\alpha-\varphi)
\cos(\beta-\varphi)
&\\
\nn
&\displaystyle
= \frac{q^2}{2} \cos(\alpha-\beta)
\int_0^{2\pi} d\varphi 
= (\hat e^{}_k\cdot \hat e^{}_t)  \frac{q^2}{2}\int_0^{2\pi} d\varphi.
&
\end{eqnarray}
Plugging this back into $(\ast)$ we see that 
\begin{eqnarray}
\nn
&\displaystyle
-\int\frac{d^2q}{(2\pi)^2}~\left[4\frac{(\hat e^{}_k\cdot q)(\hat e^{}_t\cdot q)}{q^4} - 2\frac{\hat e_t\cdot \hat e^{}_k}{q^2} \right]
&\\
\nn
&\displaystyle
= -\int\frac{d^2q}{(2\pi)^2}~\left[2\frac{\hat e^{}_k\cdot \hat e^{}_t}{q^2} - 2\frac{\hat e_t\cdot \hat e^{}_k}{q^2} \right] = 0.
&
\end{eqnarray}

\item Order $pt$: The expansion factor reads:
\begin{eqnarray}
\nn
&\displaystyle
\left.\frac{\partial^2}{\partial p\partial t} \int\frac{d^2q}{(2\pi)^2}~\frac{q^2(q-p)^2}{q^2(q-t)^2}\right|_{p,t=0} &\\
\nn
&\displaystyle
= \left.\int\frac{d^2q}{(2\pi)^2}~\frac{\partial}{\partial p}(q-p)^2\frac{\partial}{\partial t}\frac{1}{(q-t)^2} \right|_{p,t=0}
&
\\
\nn
&\displaystyle
= -4\int\frac{d^2q}{(2\pi)^2} \frac{(\hat e^{}_t\cdot q)(\hat e^{}_p\cdot q)}{q^4} 
= -2\int\frac{d^2q}{(2\pi)}\frac{\hat e^{}_t\cdot\hat e^{}_p}{q^2}.
&
\end{eqnarray}
Thus, the order $pt$ in expansion is
$$
\displaystyle -  \frac{2 (p\cdot t)}{2\pi}\ell.
$$ 

\item Order $k^2$: The corresponding factor is zero:
$$
\frac{1}{2}\frac{\partial^2}{\partial k^2}\left.\int\frac{d^2q}{(2\pi)^2}~\frac{q^2(q-k)^2}{q^2(q-k)^2}\right|_{k=0} =
\frac{1}{2}\frac{\partial^2}{\partial k^2}\int\frac{d^2q}{(2\pi)^2} = 0.
$$

\item Order $t^2$: The factor in expansion is
$$
\frac{1}{2}\frac{\partial^2}{\partial t^2}\left.\int\frac{d^2q}{(2\pi)^2} \frac{q^2}{(q-t)^2}\right|_{t=0}. 
$$
Here, it is possible to shift $q\to q+t$ and we obtain
$$ 
\frac{1}{2}\frac{\partial^2}{\partial t^2}\left.\int\frac{d^2q}{(2\pi)^2} \frac{(q+t)^2}{q^2}\right|_{t=0} =
\frac{\ell}{2\pi}.
$$

\item Order $p^2$: The corresponding expansion coefficient reads
$$
\frac{1}{2}\frac{\partial^2}{\partial p^2}\left.\int\frac{d^2q}{(2\pi)^2} \frac{q^2(q-p)^2}{q^4}\right|_{p=0} =
\frac{\ell}{2\pi}.
$$
\end{itemize}

Concluding, the leading order momentum expansion of the vertex function reads 
\begin{eqnarray}
\nn
&\displaystyle
{\rm Eq.}(\ref{eq:Expans})= \frac{(t^2-2t\cdot p + p^2)}{{D^\prime}^2}\frac{\ell}{2\pi} \rightarrow \frac{(2-1)^2}{{D^\prime}^2}\frac{\ell}{2\pi},
&
\end{eqnarray}
and the contribution to the renormalization of the interaction strength
\begin{eqnarray}
\nn
{\rm I} &\approx& -\frac{(i\lambda)^2}{{D^\prime}^2}\frac{\ell}{\pi}
\sum_i \int d1d2d3d4~ \delta(1+3-2-4)\\
\label{eq:RenInt}
& & \times
(2-1)^2\varphi^{}_{i1}\varphi^\prime_{i2}\varphi^{}_{i3}\varphi^\prime_{i4}. 
\end{eqnarray}

\subsection{Contribution II}
\begin{center}
\includegraphics[height=15mm]{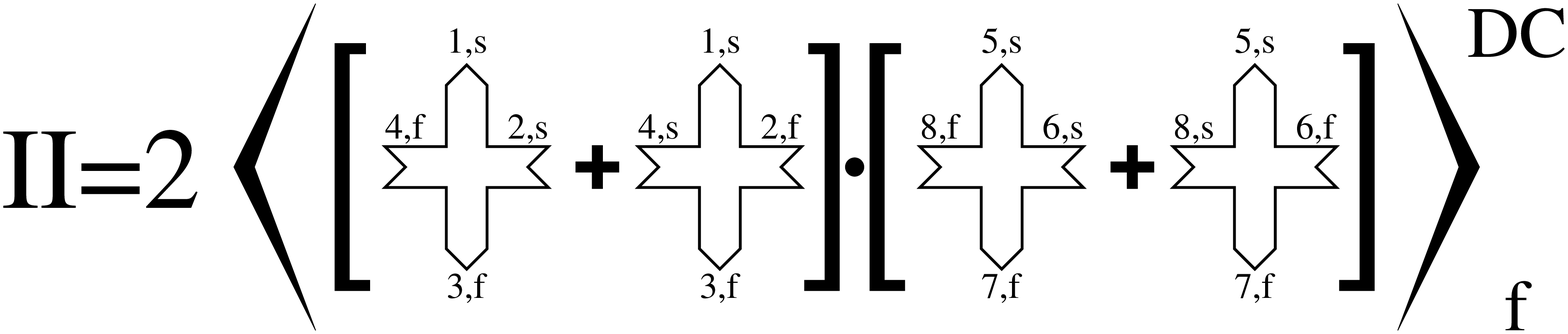}
\end{center}
Each diagram has the degeneracy one, and we may write corresponding expressions as 

\begin{eqnarray}
\nn
&\displaystyle {\rm II}  = 2(-i\lambda)^2\sum_{ij}(-1)^{i+j}\int d1d2d3d4\int d5d6d7d8 &\\
\nn
&\displaystyle 
(3-4)^2 (7-8)^2~\delta(1+3-2-4)\delta(5+7-6-8) &\\
\nn
&\displaystyle \left[\varphi^{}_{s1i}\dot\varphi^\prime_{f2i}\ddot\varphi^{}_{f3i}\varphi^\prime_{s4i}\varphi^{}_{s5j}\varphi^\prime_{s6j}\dot\varphi^{}_{f7j}\ddot\varphi^\prime_{f8j} \right. &\\
\nn
&\displaystyle
+ \varphi^{}_{s1i}\varphi^\prime_{s2i}\ddot\varphi^{}_{f3i}\dot\varphi^\prime_{f4i}\varphi^{}_{s5j}\varphi^\prime_{s6j}\dot\varphi^{}_{f7j}\ddot\varphi^\prime_{f8j} &\\
\nn
&\displaystyle 
+ \varphi^{}_{s1i}\varphi^\prime_{s2i}\dot\varphi^{}_{f3i}\ddot\varphi^\prime_{f4i}\varphi^{}_{s5j}\dot\varphi^\prime_{f6j}\ddot\varphi^{}_{f7j}\varphi^\prime_{s8j} &\\
&\displaystyle 
\left.
+ \varphi^{}_{s1i}\dot\varphi^\prime_{f2i}\ddot\varphi^{}_{f3i}\varphi^\prime_{s4i}\varphi^{}_{s5j}\ddot\varphi^\prime_{f6j}\dot\varphi^{}_{f7j}\varphi^\prime_{s8j}
\right]. &
\end{eqnarray}

After permuting fields, performing summations and integrations, and renaming variables we obtain following expressions:
\begin{eqnarray}
\nn
&\displaystyle
{\rm IIa} = 2(i\lambda)^2\sum_i\int d1d2d3d4~(3-4)^2\delta(1+3-2-4) &\\
\nn
&\displaystyle
\times\varphi^{}_{1i}\varphi^\prime_{2i}\varphi^{}_{3i}\varphi^\prime_{4i}\int d5~(5-1)^2\Pi(5)\Pi(5+2-1),&\\
\nn
&\displaystyle
{\rm IIb} = - 2(i\lambda)^2\sum_i \int d1d2d3d4~(3-4)^4\delta(1+3-2-4) &\\
\nn
&\displaystyle
\times\varphi^{}_{1i}\varphi^\prime_{2i}\varphi^{}_{3i}\varphi^\prime_{4i}
\int d5~\Pi(5)\Pi(5+3-4),&\\
\nn
&\displaystyle
{\rm IIc} =  2(i\lambda)^2\sum_i\int d1d2d3d4~(3-4)^2 \delta(1+3-2-4) &\\
\nn
&\displaystyle
\times\varphi^{}_{1i}\varphi^\prime_{2i}\varphi^{}_{3i}\varphi^\prime_{4i}\int d5~ (5-4)^2\Pi(5)\Pi(5-2-4),&\\
\nn
&\displaystyle
{\rm IId} = -2(i\lambda)^2\sum_i \int d1d2d3d4~\delta(1+3-2-4)&\\
\nn
&\displaystyle
\times\varphi^{}_{1i} \varphi^\prime_{2i} \varphi^{}_{3i} \varphi^\prime_{4i} \int d5~(5-4)^2(5-1)^2\Pi(5)\Pi(5+2-1).&
\end{eqnarray}
While evaluating contributions IIa and IIc, it suffices to take only the most divergent part from the integral over loop momentum 5 into account. This yields
\begin{eqnarray}
\nn
&\displaystyle 
{\rm IIa} = {\rm IIc} \approx 2(i\lambda)^2 \int\frac{d^2q}{(2\pi)^2}~q^2\Pi^2(q) &\\
\nn
&\displaystyle
\times\sum_i\int d1d2d3d4~(3-4)^2\delta(1+3-2-4)\varphi^{}_{1i}\varphi^\prime_{2i}\varphi^{}_{3i}\varphi^\prime_{4i} &\\
\nn
&\displaystyle
= \frac{\ell}{\pi}\frac{(i\lambda)^2}{{D^\prime}^2}\sum_i\int d1d2d3d4~\delta(1+3-2-4) &\\
\nn
&\displaystyle
\times(3-4)^2\varphi^{}_{1i}\varphi^\prime_{2i}\varphi^{}_{3i}\varphi^\prime_{4i}. &
\end{eqnarray}
For contribution Ib, it is necessary to perform a full integration over the loop momentum with help of the Feynman parametrization:
\begin{eqnarray}
\nn
\int \frac{d^2q}{(2\pi)^2} \Pi(q)\Pi(q+t) &=& \frac{1}{{D^\prime}^2} \int\frac{d^2q}{(2\pi)^2} \frac{1}{q^2(q+t)^2} \\
\nn
& = &\frac{\ell}{2\pi}\frac{1}{D^{\prime2}t^2}, 
\end{eqnarray}
where we replace momentum $(3-4)\to t$. Hence, contribution IIb reads
\begin{eqnarray}
\nn
{\rm IIb} &=& - \frac{\ell}{\pi}\frac{(i\lambda)^2}{{D^\prime}^2}~\sum_i\int d1d2d3d4~\delta(1+3-2-4) \\
\nn
&&
\times(3-4)^2\varphi^{}_{1i}\varphi^\prime_{2i}\varphi^{}_{3i}\varphi^\prime_{4i} = -{\rm IIa}.
\end{eqnarray}
The evaluation of the contribution IId goes analogously to the evaluation of the contribution I with the result

\begin{eqnarray}
\nn
{\rm IId} &=& -2\frac{(i\lambda)^2}{{D^\prime}^2}\frac{\ell}{2\pi}\sum_i\int d1d2d3d4~\delta(1+3-2-4) \\
\nn
&&
\times(2+4)^2\varphi^{}_{1i}\varphi^\prime_{2i}\varphi^{}_{3i}\varphi^\prime_{4i}.
\end{eqnarray}
An apparent problem with the momentum dependence of the vertex function can be cured if we remember that in DC limit $\varphi$ and $\varphi^\prime$ are not independent but $\varphi^{}_{-p} = \varphi^\prime_p$. Using this property we can mirror momenta $2\to -2$ and $3 \to -3$ and obtain 
\begin{eqnarray}
\nn
{\rm IId} &=& 2\frac{(i\lambda)^2}{{D^\prime}^2}\frac{\ell}{2\pi}\sum_i\int d1d2d3d4~\delta(1+2-3-4) \\
\nn
&&\times(2-4)^2 \varphi^{}_{1i}\varphi^\prime_{3i}\varphi^{}_{2i}\varphi^\prime_{4i}.
\end{eqnarray}
The sing change is due to permuting Grassmann variables. After renaming variables we get the topological structure of initial interaction term.
Summing up all contributions gives
\begin{eqnarray}
\nn
&\displaystyle
{\rm I+IIa+IIb+IIc+IId} = \frac{\ell}{\pi}\frac{(i\lambda)^2}{{D^\prime}^2}\times &\\
\nn
&\displaystyle
\sum_i\int d1d2d3d4~(3-4)^2\delta(1+3-2-4)\varphi^{}_{1i}\varphi^\prime_{2i}\varphi^{}_{3i}\varphi^\prime_{4i}.&
\end{eqnarray}
Therefore, the RG equation for the interaction strength acquires the form
\begin{equation}
i\bar\lambda = i\lambda + (-1)^{j} \frac{\ell}{\pi}\frac{(i\lambda)^2}{{D^\prime}^2}. 
\end{equation}
It is therefore convenient to distinguish between interactions in each channel:
\begin{equation}
\nn
i\bar\lambda^{}_j = i\lambda^{}_j + (-1)^{j} \frac{\ell}{\pi}\frac{(i\lambda^{}_j)^2}{{D^\prime}^2}.
\end{equation}
In continuous limit this gives Eq.~(\ref{eq:RGInt}).

\section{Scaling of the DC conductivity to one--loop order}
\label{app:cond}

The DC conductivity is calculated from Kubo formula Eq.~(\ref{eq:Kubo}) with the two-particles Green's function defined in Eqs.~(\ref{eq:2PGF}) and~(\ref{eq:AvOp}). The action is slightly changed using the acquired knowledge, as
\begin{eqnarray}
\nn
{\cal S}[\varphi] = {\cal S}^{}_0[\varphi] + {\cal S}^{}_{\rm int}[\varphi],
\end{eqnarray}
with
\begin{equation}
\nn
{\cal S}^{}_0[\varphi] = \sum_{ij=1,2}\int d1d2 ~\delta^{}_{i,j}\delta(1-2)~\varphi^{}_{i1}(i\epsilon + D^\prime \nabla^2)\varphi^\prime_{j2},
\end{equation}
and 
\begin{eqnarray}
\nn
{\cal S}^{}_{\rm int}[\varphi] &=& \sum^{}_{j=1,2} (-1)^j\lambda^{}_{j}\int d1d2d3d4~\delta(1+3-2-4) \\
\nn
&\times& (3-4)^2 \varphi^{}_{j1}\varphi^\prime_{j2}\varphi^{}_{j3}\varphi^\prime_{j4}.
\end{eqnarray}
We evaluate Eq.~(\ref{eq:PT1ord}) denoting  
\begin{eqnarray}
\nn
K^{}_0(q) &\sim& \sum^{}_{ij,p}\langle \varphi^{}_{iq}\varphi^\prime_{jp} \rangle^0,  \\
\nn
 K^{}_1(q) &\sim& \sum^{}_{ij,p}\langle \varphi^{}_{iq}\varphi^\prime_{jp}{\cal S}^{}_{\rm int} \rangle^0,
\end{eqnarray}
where
\begin{equation}
\nn
\langle \cdots \rangle^0 = \frac{1}{{\cal Z}^{}_0}\int{\cal D}[\varphi] \cdots e^{-{\cal S}^{}_0[\varphi]}.
\end{equation}
Here, the proportionality factor is $1/g$. The zeroth order contribution to the conductivity is calculated as usual:
\begin{eqnarray}
\nn
&\displaystyle
K^{}_0(q) \sim \sum^{}_{ij}\int\frac{d^2p}{(2\pi)^2} \langle \varphi^{}_{iq}\varphi^\prime_{jp} \rangle^0  &\\
\nn
&\displaystyle
= \sum^{}_{ij}\int\frac{d^2p}{(2\pi)^2} ~\frac{\delta^{}_{ij}(2\pi)^2~\delta(q-p)}{D^\prime q^2 +i \epsilon} 
= \frac{2}{D^\prime q^2 +i\epsilon},&
\end{eqnarray}
and further with $D^\prime = g/4\pi$:
\begin{equation}
\sigma^{}_0 = 2\frac{\epsilon^2}{g}\left.\frac{\partial}{\partial q^2} \frac{2}{D^\prime q^2 +i \epsilon} \right|_{q=0} = \frac{2D^\prime}{g} = \frac{1}{\pi},
\end{equation}
i.e. the usual universal DC conductivity of Dirac electron gas. The evaluation of the second term is more cumbersome. Respecting all possible (four in total) contraction combinations yields:
\begin{eqnarray}
\nn
&\displaystyle
K^{}_1(q) \sim - \sum_{ij}\int\langle \varphi^{}_{iq} \varphi^\prime_{jp}{\cal S}^{}_{\rm int}[\varphi]\rangle^0 &\\
\nn
&\displaystyle
 = \sum_{ij\alpha}(-1)^{\alpha} i\lambda^{}_\alpha\int\frac{d^2p}{(2\pi)^2} &\\
\nn
&\displaystyle
\int d1d2d3d4~(3-4)^2~\delta(1+3-2-4)&\\
\nn
&\displaystyle
\left[ 
\dot\varphi^{}_{iq} \ddot\varphi^\prime_{jp} \ddot\varphi^{}_{\alpha 1} \dot\varphi^\prime_{\alpha2} \dddot\varphi^{}_{\alpha3} \dddot\varphi^\prime_{\alpha4} + 
\dot\varphi^{}_{iq} \ddot\varphi^\prime_{jp} \ddot\varphi^{}_{\alpha 1} \dddot\varphi^\prime_{\alpha2} \dddot\varphi^{}_{\alpha3} \dot\varphi^\prime_{\alpha4} \right. &\\
\nn
&\displaystyle
\left.
+\dot\varphi^{}_{iq} \ddot\varphi^\prime_{jp} \dddot\varphi^{}_{\alpha 1} \dddot\varphi^\prime_{\alpha2} \ddot\varphi^{}_{\alpha3} \dot\varphi^\prime_{\alpha4} + 
\dot\varphi^{}_{iq} \ddot\varphi^\prime_{jp} \dddot\varphi^{}_{\alpha 1} \dot\varphi^\prime_{\alpha2} \ddot\varphi^{}_{\alpha3} \dddot\varphi^\prime_{\alpha4}
\right] &\\
\nn
&\displaystyle
= i\sum_{ij\alpha}(-1)^{\alpha} \lambda^{}_\alpha \delta^{}_{i\alpha}\delta^{}_{j\alpha}\delta^{}_{\alpha\alpha} 
& \\
\nn
&\displaystyle
\int\frac{d^2p}{(2\pi)^2} \int d1d2d3d4~(3-4)^2~\delta(1+3-2-4) & \\
\nn
&\displaystyle \left[ 
-\delta(2-q)\delta(1-p)\delta(3-4)K^{}_0(q)K^{}_0(p)K^{}_0(3) \right.&\\
\nn
& +  \delta(4-q)\delta(1-p)\delta(3-2)K^{}_0(q)K^{}_0(p)K^{}_0(3)  &\\
\nn
&\displaystyle   
-\delta(4-q)\delta(3-p)\delta(1-2)K^{}_0(q)K^{}_0(p)K^{}_0(2) &\\
\nn
&\displaystyle
\left.+ \delta(q-2) \delta(3-p) \delta(1-4) K^{}_0(q)K^{}_0(p)K^{}_0(4)
\right]. 
&
\end{eqnarray}
While contributions from the first and third terms vanish, both other give equal finite contributions:
\begin{equation}
K^{}_1(q) \sim 2i \sum_{j=1,2} \frac{\displaystyle \left[(-1)^j \lambda^{}_j\right]}{\displaystyle(D^\prime q^2+i\epsilon)^2} \int \frac{d^2k}{(2\pi)^2}~\frac{k^2+q^2}{D^\prime k^2 + i\epsilon}.
\end{equation}
The integral
$$
\int\frac{d^2k}{(2\pi)^2}~\frac{k^2}{D^\prime k^2+i\epsilon} 
$$
behaves well in DC limit ($\epsilon\to0$) as it does not develop any IR divergences. The corresponding contribution disappears in continuous limit and does not affect the conductivity at large scales. Therefore, the main contribution  arises from the following expression:
\begin{equation}
\nn
K^{}_1(q) \sim \frac{2iq^2}{(i\epsilon)^2} \sum_{j=1,2}\int\frac{d^2k}{(2\pi)^2}~\frac{\left[(-1)^j\lambda^{}_j\right]}{D^\prime k^2+i\epsilon}. 
\end{equation}
In DC limit, the integral diverges logarithmically. The corresponding conductivity correction reads
\begin{equation}
\nn
\sigma^{}_1 = 2{\epsilon^2} \left.\frac{\partial}{\partial q^2} K^{}_1(q)\right|_{q=0} = 
- 2i\frac{\sigma^{}_0}{gD^\prime}\sum^{}_{j=1,2}\left[(-1)^j\lambda^{}_j\right]\ell,
\end{equation}
and the full expression for the renormalized conductivity:
\begin{equation}
\bar\sigma = \sigma^{}_0 - 2i \frac{\sigma^{}_0}{gD^\prime} \sum_{j=1,2} \left[(-1)^j \lambda^{}_j\right]\ell. 
\end{equation}
Continuous limit of this expression yields Eq.~(\ref{eq:CondRgEq}).

\end{document}